\documentclass[sigconf,9pt]{acmart}

\usepackage{amsmath}
	\usepackage{cancel}
\usepackage{mathtools}
\usepackage{graphicx} 
\usepackage{url}
\usepackage{enumitem} 

\usepackage{stmaryrd} 

\usepackage{upgreek} 
\usepackage{bbding} 

\usepackage{listings} 
\usepackage{tabularx} 
\usepackage{booktabs} 
\usepackage{multirow}
\usepackage{hhline} 
\usepackage{diagbox}
\usepackage{pgfplotstable} 

\usepackage{xcolor,colortbl}    

\usepackage{tikz} 
\usetikzlibrary{matrix}
\usetikzlibrary{calc}
\usetikzlibrary{math}
\usetikzlibrary{arrows.meta,positioning}
\usetikzlibrary{intersections, pgfplots.fillbetween}

\usepackage{easybmat} 

\usepackage{adjustbox}
\def\eox{\unskip\kern 10pt{\unitlength1pt\linethickness{.4pt}$\diamondsuit${}}} 

\usepackage{rotating}		

\usepackage{xspace} 

\usepackage{subcaption} 
\newcommand{\revision}[1]{{\color{black}{#1}}}

\newcommand{\hide}[1]{}

\usepackage[ruled,noend,linesnumbered]{algorithm2e} 

\usepackage{setspace} 

\DontPrintSemicolon     
\SetNlSty{}{}{}                
\SetAlgoInsideSkip{smallskip}   

\SetAlFnt{\small}			
\SetAlCapFnt{\small}		
\SetAlCapNameFnt{\small}

\usepackage[capitalise,nameinlink]{cleveref} 

\crefname{section}{Sec.}{Secs.}
\crefname{example}{Ex.}{Exes.}

\hypersetup{                                                            
	pdfpagemode=UseNone,               
    pdfstartview=Fit,
    pdfpagelayout=SinglePage,       
    bookmarks=false,
    bookmarksnumbered=true,
    bookmarksopen=false,             
    pdftex=true,
    unicode,
    colorlinks=true,       
    linkcolor=magenta,          
    citecolor=cyan,  
    filecolor=magenta,      
    urlcolor=blue           
}

\usepackage{aliascnt}  	

\newaliascnt{corollary}{theorem}

\aliascntresetthe{corollary}

\newaliascnt{example}{theorem}
\newtheorem{example}[example]{Example}
\aliascntresetthe{example}

\newaliascnt{definition}{theorem}

\aliascntresetthe{definition}

\newaliascnt{proposition}{theorem}

\aliascntresetthe{proposition}

\newaliascnt{lemma}{theorem}

\aliascntresetthe{lemma}

\newaliascnt{conjecture}{theorem}

\aliascntresetthe{conjecture}

\newtheorem{questionW}{Question}
\newtheorem{resultW}{Result}

{\end{itshape}
}
\AtEndPreamble{%
    \hypersetup{colorlinks,
      linkcolor=purple,
      citecolor=blue,
      urlcolor=ACMDarkBlue,
      filecolor=ACMDarkBlue}}

\usepackage{tcolorbox}		
\tcbuselibrary{breakable,skins}		

\tcbset{examplestyle/.style={
		enhanced jigsaw,	
		colback=blue!08,	
		colframe=blue!08,	
		arc=2mm,
		boxrule=0pt,		
		left=1mm,
		right=1mm,
		left skip= 0mm,  
		right skip= 0mm, 
		top=1mm,		
		bottom=1mm,		
		breakable,		
		parbox = false,		
		before={\par\pagebreak[0]\vspace{1mm}\parindent=0pt},		
		after={\par\pagebreak[0]\vspace{1mm}\parindent=0pt},				
		bottomrule = 0mm,
		boxsep = 0mm,					
		topsep at break=0pt,			
		bottomsep at break=0pt,			
		pad at break=0mm,
		pad before break=1mm,		
		pad after break=1mm,		
		bottomrule at break=0mm,
		toprule at break=0mm,		
		}}

\usepackage{footnote}

\tcolorboxenvironment{example}{examplestyle}

\makeatletter
\DeclareRobustCommand*\uell{\mathpalette\@uell\relax}
\newcommand*\@uell[2]{
  \setbox0=\hbox{$#1\ell$}
  \setbox1=\hbox{\rotatebox{10}{$#1\ell$}}
  \dimen0=\wd0 \advance\dimen0 by -\wd1 \divide\dimen0 by 2
  \mathord{\lower 0.1ex \hbox{\kern\dimen0\unhbox1\kern\dimen0}}
}
\makeatother

\usepackage{marginnote}

\newcommand{\introparagraph}[1]{\textbf{#1.}} 

\renewcommand{\epsilon}{\varepsilon} 

\usepackage{scalerel}
\usepackage{tikz}
\usetikzlibrary{svg.path}

\definecolor{orcidlogocol}{HTML}{A6CE39}
\tikzset{
  orcidlogo/.pic={
    \fill[orcidlogocol] svg{M256,128c0,70.7-57.3,128-128,128C57.3,256,0,198.7,0,128C0,57.3,57.3,0,128,0C198.7,0,256,57.3,256,128z};
    \fill[white] svg{M86.3,186.2H70.9V79.1h15.4v48.4V186.2z}
                 svg{M108.9,79.1h41.6c39.6,0,57,28.3,57,53.6c0,27.5-21.5,53.6-56.8,53.6h-41.8V79.1z M124.3,172.4h24.5c34.9,0,42.9-26.5,42.9-39.7c0-21.5-13.7-39.7-43.7-39.7h-23.7V172.4z}
                 svg{M88.7,56.8c0,5.5-4.5,10.1-10.1,10.1c-5.6,0-10.1-4.6-10.1-10.1c0-5.6,4.5-10.1,10.1-10.1C84.2,46.7,88.7,51.3,88.7,56.8z};
  }
}

\RequirePackage{etoolbox}
\DeclareRobustCommand\orcidicon[1]{\href{https://orcid.org/#1}{\mbox{\scalerel*{
\begin{tikzpicture}[yscale=-1,transform shape]
\pic{orcidlogo};
\end{tikzpicture}
}{|}}}} 
\usepackage{array}
\usepackage{makecell}

\AtBeginDocument{%
  }

\copyrightyear{2023}
\acmYear{2023}
\setcopyright{acmlicensed}\acmConference[SIGMOD-Companion
'23]{Companion of the 2023 International Conference on Management of
Data}{June 18--23, 2023}{Seattle, WA, USA}
\acmBooktitle{Companion of the 2023 International Conference on
Management of Data (SIGMOD-Companion '23), June 18--23, 2023, Seattle,
WA, USA}
\acmPrice{15.00}
\acmDOI{10.1145/3555041.3589732}
\acmISBN{978-1-4503-9507-6/23/06}

\begin{document}

\title{DIALITE: Discover, Align and Integrate Open Data Tables}


\author{Aamod Khatiwada}
\affiliation{%
  \institution{Northeastern University}
  \city{Boston}
  \state{Massachusetts, USA}
}
\email{khatiwada.a@northeastern.edu}

\author{Roee Shraga}
\affiliation{%
  \institution{Northeastern University}
  \city{Boston}
  \state{Massachusetts, USA}
}
\email{r.shraga@northeastern.edu}

\author{Ren\'ee J. Miller}
\affiliation{%
  \institution{Northeastern University}
  \city{Boston}
  \state{Massachusetts, USA}
  }
\email{miller@northeastern.edu}
\renewcommand{\shortauthors}{Aamod Khatiwada et al.}

\begin{abstract}
We demonstrate a novel table discovery pipeline called DIALITE that allows users to discover, integrate and analyze open data tables. DIALITE has three main stages. First, it allows users to discover tables from open data platforms using state-of-the-art table discovery techniques. Second, DIALITE integrates the discovered tables to produce an integrated table. Finally, it allows users to analyze the integration result by applying different downstreaming tasks over it. Our pipeline is flexible such that the user can easily add and compare additional discovery and integration algorithms.
\end{abstract}



\maketitle
\section{introduction}
\label{section:introduction}
Data discovery has become an important component in data science pipeline.
The discovery process uses techniques such as keyword search~\cite{2020_shraga_table_retrieval} 
and table search~\cite{2018_nargesian_tus, 2023_khatiwada_santos, DBLP:journals/corr/abs-2210-01922} 
to discover a set of tables (datasets). 
Data scientists use such tables to support decision-making processes, train machine learning models, perform statistical analysis, and so on. After discovery, a natural step is to integrate the discovered tables. An integrated table provides a unified view of the data and allows users to run queries and analyses that go beyond a single table. 

While integrating tables, we intend to combine tuples from different tables in a maximal way such that the integrated tuples carry as much information as possible. This enriches the analysis and decision-making process after integration and also improves the quality of downstream applications. The widely known outer-join~\cite{cowbook} 
operator is not associative and does not aim to maximize the connections among the integrated tuples
~\cite{1994_legaria_outerjoin_as_disjunctions}. Accordingly, \textit{Full Disjunction (FD)}~\cite{1994_legaria_outerjoin_as_disjunctions} has been understood as a natural way of assembling partial pieces of information (facts) such that it maximizes the connections among these facts~\cite{1996_ullman_outer_join_gamma_cycles}. FD can be viewed as an associative version of outer join~\cite{2006_cohen_poly_delay_fD} and has been used to integrate information across relational tables~\cite{2006_cohen_poly_delay_fD, 2019_paganelli_parallelizing_fd} and web tables~\cite{2019_paganelli_parallelizing_fd}. 
In a recent paper~\cite{2022_khatiwada_alite}, we proposed ALITE, a new algorithm based on Full Disjunction to integrate  
tables discovered in data lakes. 
ALITE was shown to be correct and faster than the existing FD algorithms~\cite{2006_cohen_poly_delay_fD, 2019_paganelli_parallelizing_fd} while integrating real open data lake tables in practice~\cite{2022_khatiwada_alite}. Additional details on other integration operators and their comparison against FD are available in the full ALITE paper~\cite{2022_khatiwada_alite} where FD was shown to be a better semantics for integration because it produces a result over which a downstream task like entity-resolution performs more accurately.

In this demonstration, we propose DIALITE, a novel system that discovers, aligns, and integrates open data tables. It extends the aforementioned ALITE~\cite{2022_khatiwada_alite}, which does not perform discovery but instead takes a set of tables as input, aligns the matching columns using holistic schema matching, and applies the FD to get an integrated table. 
DIALITE offers state-of-the-art systems for different table discovery tasks~\cite{2019_zhu_josie, 2023_khatiwada_santos} before applying ALITE to integrate them. Furthermore, DIALITE also offers new downstream analytics 
(that have not been previously considered) to evaluate the quality of integration. Specifically, DIALITE allows users to 

\begin{enumerate}
    \item Upload a table or randomly generate one using GPT-3 as a query and discover related tables from a given data lake that are unionable~\cite{2023_khatiwada_santos} or joinable~\cite{2019_zhu_josie} 
    with the query table. Apart from the available table discovery algorithms, DIALITE also allows users to add new algorithms for table search based on their preference.
    \item Integrate the discovered tables (or upload a set of tables to be integrated) using a novel table integration system called ALITE~\cite{2022_khatiwada_alite}. Besides ALITE, we allow users to add new integration operators 
    within our extendible architecture.
    \item Analyze and compare the table integrated using ALITE 
    against alternative integration techniques by performing downstream applications. In our demo, we consider outer join as an alternative integration technique (or other queries/methods added by the user) and present 
    data analytics (common aggregations and statistics) and entity-resolution as downstream applications.
    Both, when applied over real tables (which can be incomplete) will show the dramatic difference between maximally integrating information using FD vs. using outer joins.
\end{enumerate}

\introparagraph{Related Work}
To the best of our knowledge, DIALITE is the first system that enables a full table search pipeline starting from table discovery, followed by table integration and downstream applications over the integrated table. An earlier system called COCOA~\cite{2021_esmailoghli_cocoa}
focuses on finding joinable tables with correlated attributes that expand the query table. After finding the tables, COCOA applies LEFT JOIN between the query table and the discovered table as the integration operator. DIALITE, on the other hand, finds a set of  related tables that can be integrated with the query table (including unionable and joinable tables). 
Auctus~\cite{2021_castelo_auctus} also discovers related tables and integrates table pairs by applying inner join or union. Unlike Auctus, DIALITE does not limit the number of tables to be integrated.
In addition, DIALITE uses FD to integrate the tables, which has been shown in theory to maximize the connections among the tuples in the tables~\cite{1996_ullman_outer_join_gamma_cycles}. 

Existing table discovery techniques search for joinable, unionable, or related tables. \textbf{JOSIE}~\cite{2019_zhu_josie} and \textbf{LSH Ensemble}~\cite{DBLP:journals/pvldb/ZhuNPM16}, for instance, take a query table as input with a query column marked by the user and returns a set of top-k tables that are joinable with the query table. 
\textbf{SANTOS}~\cite{2023_khatiwada_santos} takes a query table and 
discovers a set of top-k semantically unionable tables as output.
Other table search techniques~\cite{2018_nargesian_tus, DBLP:journals/corr/abs-2210-01922}
 also focus on table discovery from large repositories. But the important question of how to integrate the discovered tables is not addressed.
 \textbf{ALITE}~\cite{2022_khatiwada_alite} provides a solution to this question. As the data lake tables may lack consistent and meaningful column headers,
 ALITE applies holistic schema matching over the set of searched tables and assigns a dummy column header called an Integration ID to the set of matching columns. Then, it applies a natural FD over the integration IDs to integrate the tables. The integration process outputs an integrated table with 
maximally integrated tuples~\cite{1996_ullman_outer_join_gamma_cycles,1994_legaria_outerjoin_as_disjunctions}.
\section{System Description}
\label{section:system-description}
\begin{figure}[ht]
  \includegraphics[scale = 0.25]{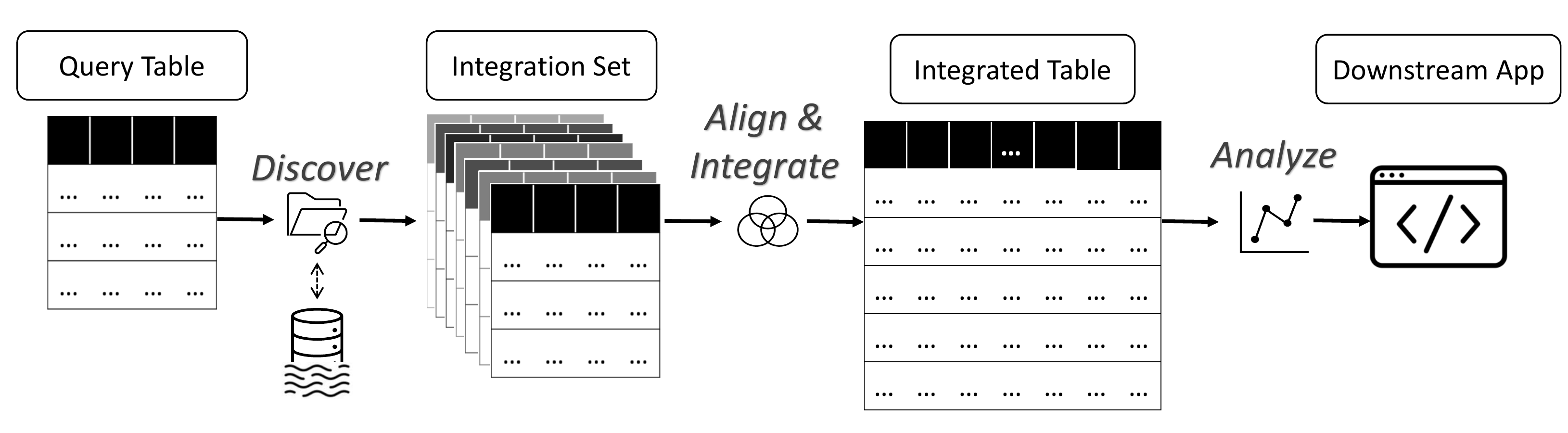}
  \caption{An overview of DIALITE.}
\label{fig:dealite}
\end{figure}

DIALITE, outlined in \cref{fig:dealite}, has three stages (discover, align \& integrate, and analyze) which are included in our demo plan (\cref{section:demonstration-plan}). 

DIALITE offers various options for discovery, alignment, integration, and analysis, including the ability for users to implement their own methods.
The pipeline begins with a user-provided query table 
$Q$. However, we also allow the user to randomly generate a table using GPT-3~\cite{2020_floridi_gpt3}. We now detail the components.

\subsection{Discover}\label{sec:discover}
Given a query table $Q$, DIALITE uses a data discovery method to find tables in a data lake (table repository) $\mathcal{D}$ that are unionable~\cite{2023_khatiwada_santos}, joinable~\cite{2019_zhu_josie}, or simply semantically similar. DIALITE allows the user to choose among existing discovery algorithms including SANTOS~\cite{2023_khatiwada_santos} and LSH Ensemble~\cite{DBLP:journals/pvldb/ZhuNPM16}. Alternatively, a user can use their own discovery algorithm by implementing a similarity method between two tables. As we apply ALITE~\cite{2022_khatiwada_alite} for alignment and integration, the discovery phase is agnostic to the type of search.
ALITE, aiming to maximize the connections among the tables, decides how the tables should be integrated. The output of discovery is an \emph{integration set} $D\subseteq\mathcal{D}$, a set of tables to be integrated, including the query table. 
\revision{
Our discovery techniques allow users to control the number of tables returned, 
and our system permits users to select a subset of the discovered tables to be integrated.}

\subsection{Align and Integrate}\label{sec:alignintegrate}
Given an integration set $D$, DIALITE uses ALITE~\cite{2022_khatiwada_alite} to integrate the tables.
The integration set can be derived from the discovery part (\cref{sec:discover}) or provided as input by the user. The latter represents a traditional data integration scenario where the integration set is given. 
As shown in \cref{fig:dealite}, this stage outputs an integrated table.

ALITE is composed of two main parts, namely align and integrate.
Note that we do not require reliable column headers in the integration set. The first part of ALITE (Align) applies holistic schema matching to identify common columns in the integration set.
The matched columns are given the same integration ID. Using these ids as attribute names, the integrate part applies (natural) FD. Specifically, ALITE uses a new holistic schema matching algorithm that was shown to outperform state-of-the-art matchers.  
ALITE uses a novel algorithm to compute the FD shown to be correct and efficient on real tables (including tables having nulls)~\cite{2022_khatiwada_alite}.
DIALITE also allows users to add alternative integration operators, e.g., outer join, which is included for demonstration. 

\subsection{Analyze}\label{sec:analyze}
Given an integrated table (that can also be uploaded as a file by the user), DIALITE allows the user to explore the benefits of integration. Specifically, the user can choose a downstream application
to apply over the integrated table. A simple application is an aggregation query that can be applied over the integrated table as we illustrate in \cref{sec:demo_plan1}. In \cref{sec:demo_plan2} we also present a more complex downstream application, entity-resolution (ER). 

\subsection{Implementation}
\revision{DIALITE is implemented in Python 3.8 and the demonstration uses a web application.}\footnote{\label{footnote:gitlink}\url{https://github.com/northeastern-datalab/dialite}}
Within our pipeline, we use SANTOS
\footnote{\url{https://github.com/northeastern-datalab/santos}}, LSH Ensemble
\footnote{\url{https://github.com/ekzhu/datasketch}}
and ALITE \footnote{\url{https://github.com/northeastern-datalab/alite}} using their publicly available code. Also, we use the
\textit{py\_entitymatching} package to show ER as a downstream application.\footnote{\label{footnote:er}\url{https://github.com/anhaidgroup/py_entitymatching}}
\revision{ 
We allow users to interact with the system after each step so that they can validate the intermediate results.} 
\section{Demonstration Plan}
\label{section:demonstration-plan}

Next, we illustrate our demonstration. 
The link to a demonstration video is available in the github repository.$^{\ref{footnote:gitlink}}$ 
Our demo contains two parts. First, we present the use case of the pipeline as described in~\cref{section:system-description}. This part is actually composed of three demonstration items, that can be demonstrated independently. Then, we demonstrate DIALITE's extensibility to new algorithms for discovery, integration, and analysis. 

\subsection{DIALITE Use case} \label{sec:demo_plan1}
\introparagraph{Discovery}
The users of DIALITE would be able to upload a query table in CSV format from an existing pool of tables.\footnote{For the demonstration itself, we also allow users to randomly generate a query table (see \cref{sec:demo_plan2}). Also, a user may choose to upload their query table noting that it may be off-topic wrt the data lake, which may yield no results.}  
Note that we will provide a data lake for the users to use in the demonstration. The tables in this data lake are real tables from open data and are currently preprocessed for our discovery algorithms~\cite{2023_khatiwada_santos, DBLP:journals/pvldb/ZhuNPM16}. Specifically, the indexes used in SANTOS~\cite{2023_khatiwada_santos}  and LSH Ensemble~\cite{DBLP:journals/pvldb/ZhuNPM16}  are built offline, i.e., they are already available for the user to use. The users can easily preprocess and link their own data lake.

The set of discovered tables by all the table discovery systems is stored for the next step. As there may be an overlap in unionable and joinable search results, we persist \emph{the set of tables} found by all techniques to form an integration set.

\begin{figure}[th]
  \includegraphics[scale = 0.44]{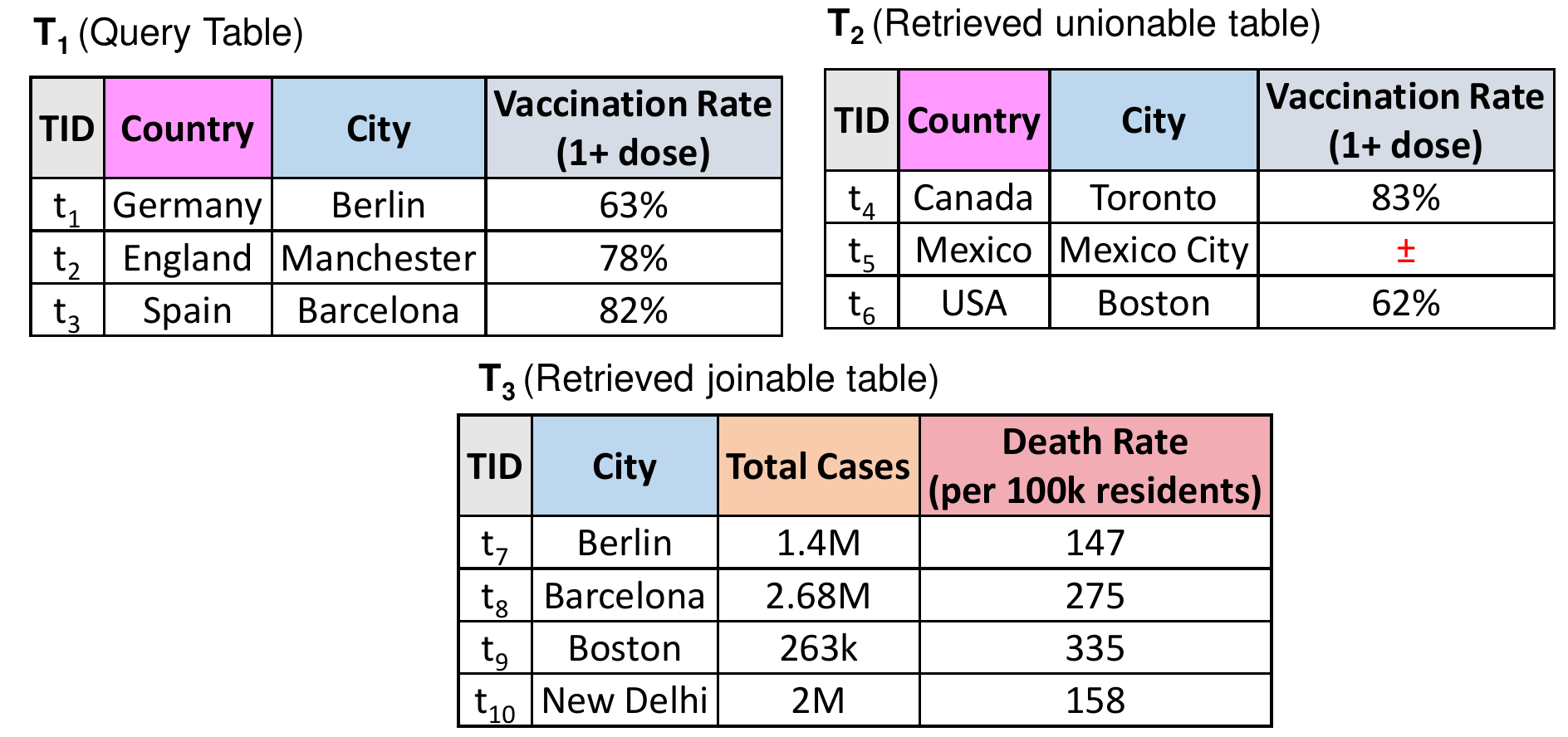}
  \caption{
  Tables detailing COVID-19 cases in different places. The symbol $\pm$ represents null values present in the input tables (``\textit{missing nulls}'').}
\label{fig:demo_input_table}
\end{figure}

\begin{example}
\label{ex:discover_tables}
Consider tables $T_1$, $T_2$ and $T_3$ about COVID-19 cases in different places shown in~\cref{fig:demo_input_table}. Here, TID (Tuple ID) is not a real data column and it is added only to refer to the tuples. Let, $T_1$ be the query table, and tables $T_2$ and $T_3$ reside in the data lake repository. 
\revision{In our demo, a user selects \texttt{City} as an intent column and query column to search for a unionable table using SANTOS~\cite{2023_khatiwada_santos} and a joinable table using LSH Ensemble~\cite{DBLP:journals/pvldb/ZhuNPM16} respectively. Please see respective papers for additional details on the search algorithms.}
Let, $T_2$ and $T_3$ be results of the unionable and joinable searches respectively. So the result of the discovery step is an integration set of tables $T_1$, $T_2$, and $T_3$. Note that the names of columns are presented for simplicity and are not used by the discovery techinques, which are designed for the ambiguty of data lakes, i.e., unreliable/missing metadata.
\end{example}

\introparagraph{Align and Integrate}
Once the integration set is formed, DIALITE allows user to apply ALITE's holistic schema matching to generate integration IDs. 
Over such IDs, we apply ALITE's FD.

\begin{figure}[th]
  \includegraphics[scale = 0.46]{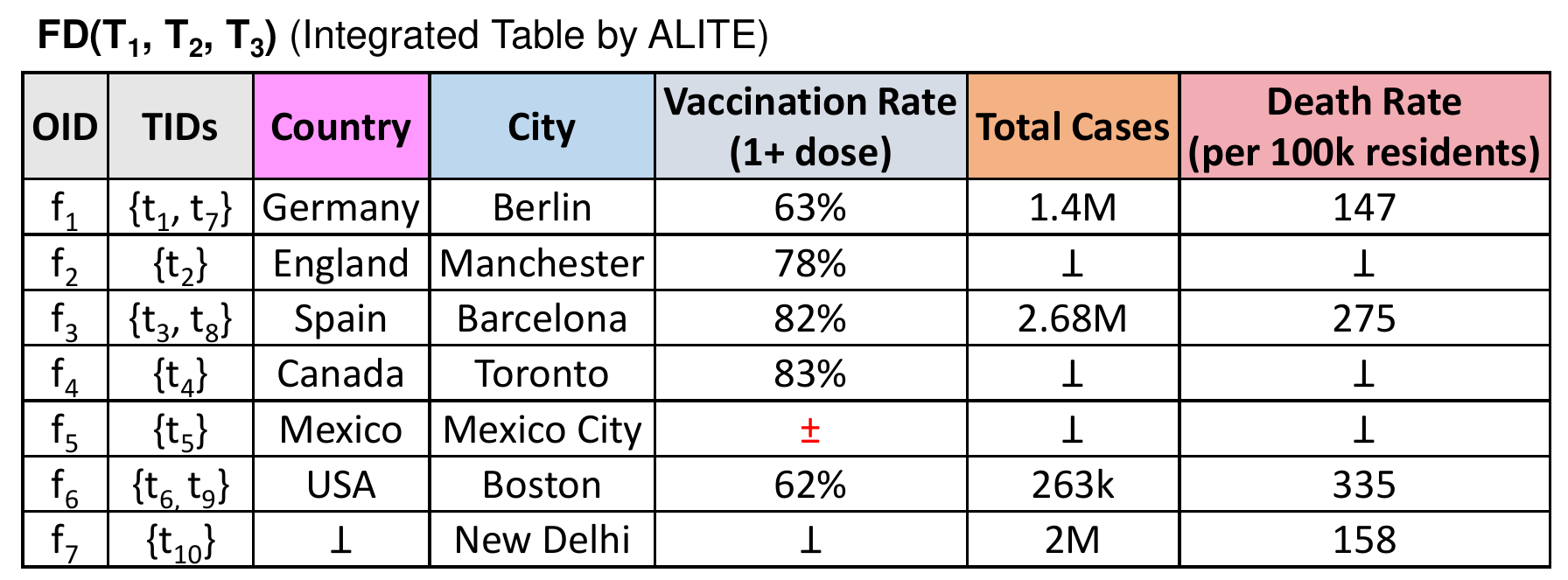}
  \caption{Result of applying ALITE over the tables in~\cref{fig:demo_input_table}. The symbol $\perp$ represents the null values produced during the integration due to missing information (``\textit{produced nulls}'').}
\label{fig:demo_output_table}
\end{figure}

\begin{example}
\label{ex:align_and_integrate}
Consider the integration set of tables $T_1$, $T_2$ and $T_3$ shown in~\cref{fig:demo_input_table} is formed after table discovery step as illustrated in~\cref{ex:discover_tables}. DIALITE applies ALITE's integration algorithm over these tables that returns an integrated table as shown in~\cref{fig:demo_output_table}. The integration semantics is explained in ALITE paper~\cite{2022_khatiwada_alite}.
\end{example}

\introparagraph{Analyze}
After integration, the next feature that ALITE offers is the analysis of the integrated table.
We show the use of an aggregation query over the integrated table.
\begin{example}
With the integrated table, we now allow the user to use queries that go beyond the single tables. For example, over the integrated table~(\cref{fig:demo_output_table}), the user can find that Boston is the city with the lowest vaccination rate and Toronto has the highest. Trying to understand the reason for that, the user may explore the relationship between vaccination rates (given in $T_1$ and $T_2$), number of cases and death rates (given in $T_3$). For example, the user can compute the correlation between vaccination and death rates that shows a positive (pearson) correlation of $0.16$ and (somewhat surprising) correlation of $0.9$ between case numbers and vaccination rates. While a bit counter-intuitive, the analysis reveals an interesting insight about the nature of vaccinations, suggesting that in cities with higher death rates and more cases, the government is focusing on vaccination programs and the people are more willing to vaccinate.
\end{example}

\subsection{DIALITE Extendibility}\label{sec:demo_plan2}

As described in \cref{section:system-description}, the user can extend DIALITE by implementing their own alternative components in the pipeline. Specifically, we aim to demonstrate the ability of users to implement (using python code) new discovery algorithms, integration methods and perform their required analysis. 

\begin{figure}[h]
  \includegraphics[scale = 0.5]{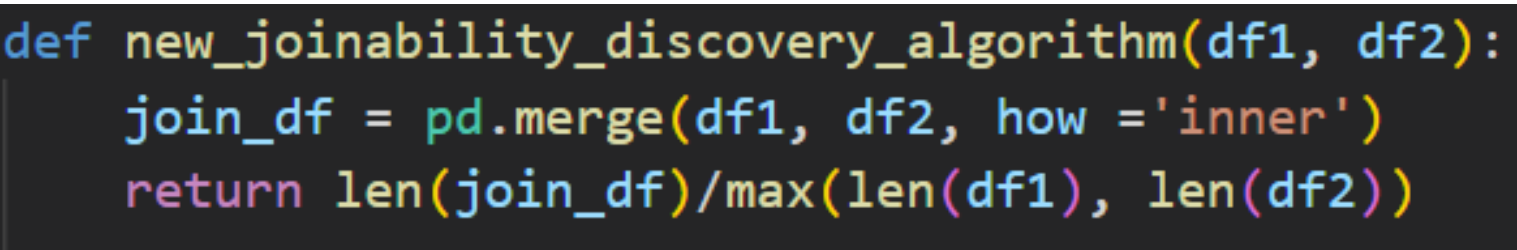}
  \caption{Implementing user-defined discovery algorithm based on inner join.}
\label{fig:new_discovery_alg}
\end{figure}

\begin{example}
    For illustration, a sample code snippet to add a new joinable table discovery algorithm is provided in~\cref{fig:new_discovery_alg}. The user basically implements a similarity function between two datasets (df1 and df2) that is used by DIALITE for table discovery.
\end{example}

\begin{figure}[h]
  \includegraphics[scale = 0.35]{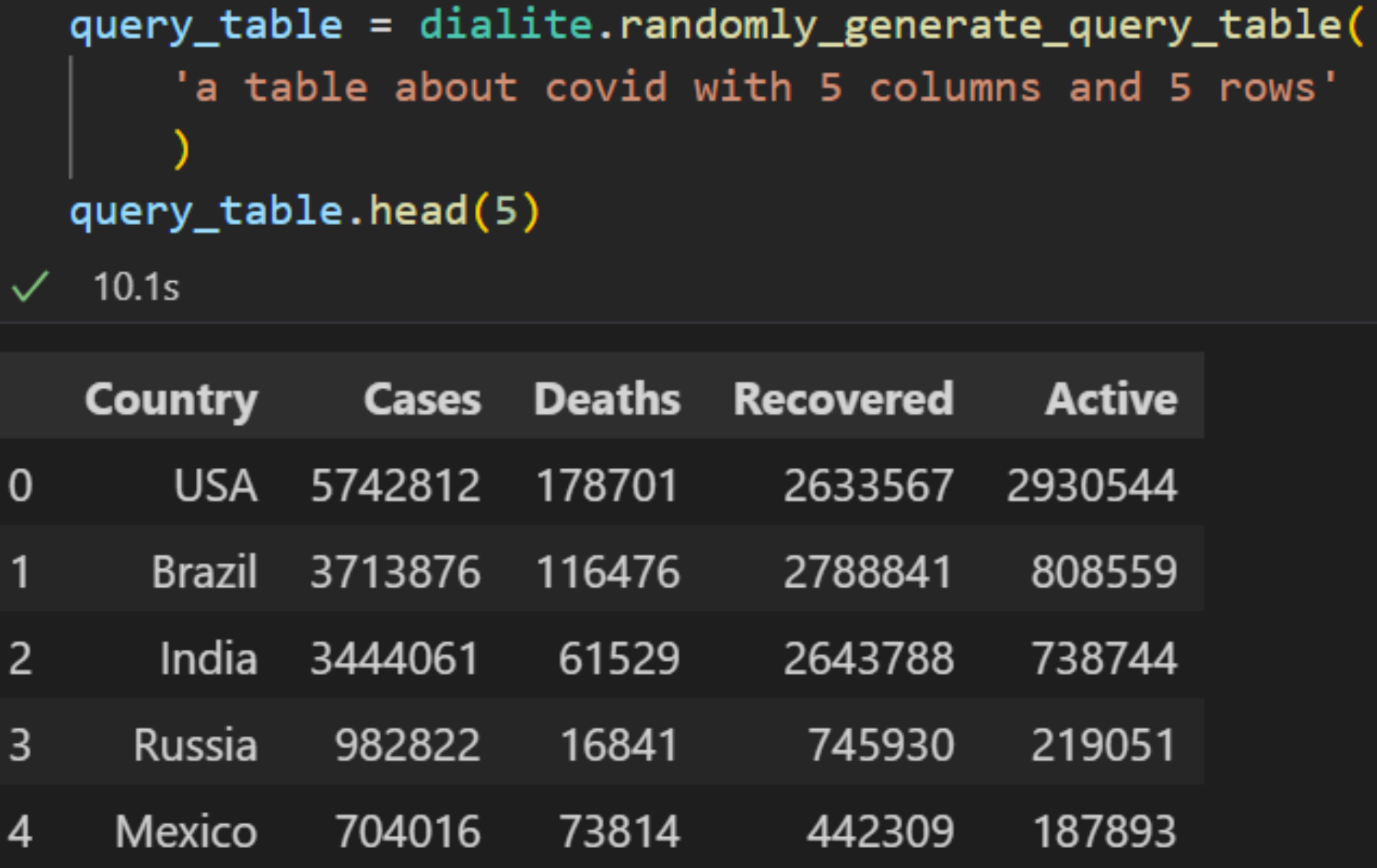}
  \caption{A code snippet to generate query table using GPT-3.}
\label{fig:query_using_gpt}
\end{figure}

We also consider a scenario where the user may not have a query table to start the analysis. So, DIALITE allows users to use simple prompts to generate the query table for the analysis. We demonstrate this feature using GPT-3 based implementation~\cite{2020_floridi_gpt3} and allow the user to generate a query table based on prompt as illustrated in~\cref{fig:query_using_gpt}. Here, we generate a query table about COVID-19 cases that has 5 columns and 5 rows. 

\begin{figure}[h]
  \includegraphics[scale = 0.35]{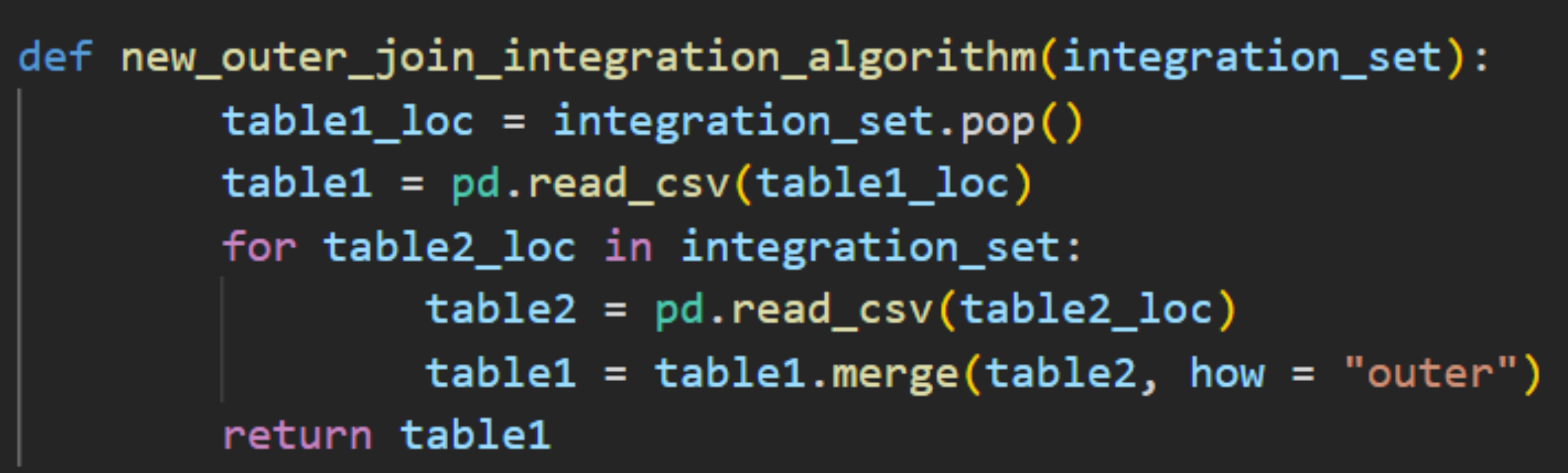}
  \caption{Implementing outer join as an integration operator.}
\label{fig:outer_join_alg}
\end{figure}

Furthermore, the users can add an alternative integration operator over the default integration system (ALITE). For example, \cref{fig:outer_join_alg} shows a user-defined code snippet that implements the commonly used outer-join operator for integration. In the demonstration, we illustrate the benefit of using ALITE over the standard outer join as shown in the following example.

\begin{figure}[t]
  \includegraphics[scale = 0.43]{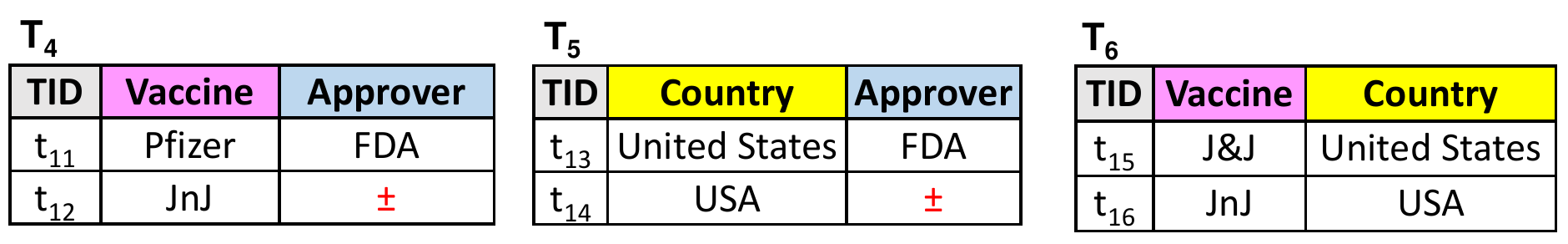}
  \caption{An integration set of Tables about COVID-19 vaccines, their country of origin and their approvers.}
\label{fig:er_input_tables}
\end{figure}

\begin{figure}[t]
  \includegraphics[scale = 0.45]{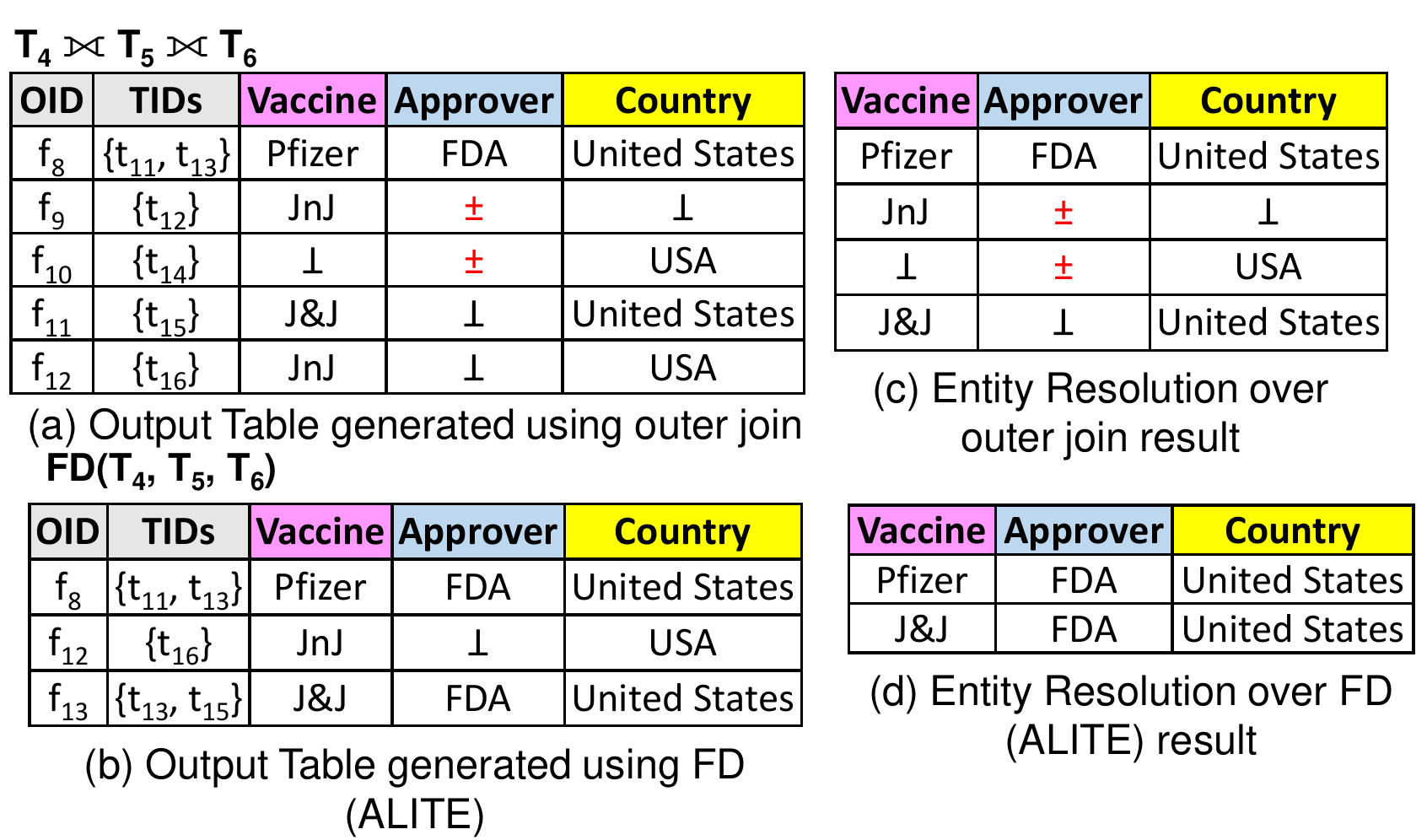}
  \caption{Integrating tables in~\cref{fig:er_input_tables} using outer join and FD.}
\label{fig:er_output_tables}
\end{figure}

\begin{example}
\label{ex:entity_resolution_example}
Consider tables $T_4$, $T_5$ and $T_6$ shown in \cref{fig:er_input_tables} (a) forms an integration set after table discovery where, the tables describe the COVID-19 vaccines, their country of origin, and the regulatory agency that approved the vaccines. For sake of illustration, assume that a user used outer-join as an alternative integration algorithm (see \cref{fig:outer_join_alg}). The results of applying outer-join and ALITE (FD) over these tables are shown in \cref{fig:er_output_tables} (a) and \cref{fig:er_output_tables} (b) respectively. 
Now let us assume that the user wants to apply Entity Resolution (ER) as a downstream application by applying 
\textit{py\_entitymatching}.$^{\ref{footnote:er}}$ 
This analysis over outer join and FD results are shown in \cref{fig:er_output_tables}(c) and \cref{fig:er_output_tables}(d), respectively. Outer join produces more output tuples than FD; yet, it does not produce any tuple containing the agency that approved the Johnson \& Johnson (J\&J) vaccine.
FD, on the other hand, produces an output tuple $f_{13}$ that provides this information (as it can be produced using $t_{13}$ and $t_{15}$). Furthermore, since outer join produces incomplete tuples, ER can not resolve $f_{9}$ and $f_{10}$. This shows an advantage of using the default ALITE operator instead of outer join to integrate the tables.
\end{example}
\begin{acks}
This work was supported in part by NSF
under award numbers IIS-1956096 and IIS-2107248.
\end{acks}
\bibliographystyle{ACM-Reference-Format}
\bibliography{main}
\end{document}